\documentclass[twocolumn,amsmath,amssymb,groupaddress,reprint]{revtex4-1}
\usepackage{graphicx}
\usepackage[dvipdfmx]{color}
\usepackage{dcolumn}
\usepackage{bm}
\usepackage{txfonts}
\usepackage[utf8]{inputenc}
\usepackage[T1]{fontenc}
\usepackage{mathptmx}
\usepackage{etoolbox}
\usepackage{multirow}
\usepackage[hypertex,colorlinks=true,linkcolor=black,citecolor=blue,filecolor=blue,urlcolor=blue,setpagesize=false,nesting=true]{hyperref}

\makeatletter
\def\@email#1#2{%
 \endgroup
 \patchcmd{\titleblock@produce}
  {\frontmatter@RRAPformat}
  {\frontmatter@RRAPformat{\produce@RRAP{*#1\href{mailto:#2}{#2}}}\frontmatter@RRAPformat}
  {}{}
}%
\makeatother

\begin{document}
\title{An empirical model for deriving information on ion dynamics from $\mu$SR data\\ via the conventional Kubo-Toyabe function analysis}

\author{Ryosuke Kadono}
\affiliation{Muon Science Laboratory, Institute of Materials Structure Science, \\High Energy Accelerator Research Organization (KEK), Tsukuba, Ibaraki 305-0801, Japan}

\begin{abstract}
We present an empirical model that analytically relates the parameters of the extended Kubo-Toyabe (KT) function $G_z^{\rm EA}(t;\Delta,\nu_{\rm i},Q)$, which describes muon spin relaxation ($\mu$SR) induced by jump diffusion of ions using the Edwards-Anderson parameter $Q$, to those of the conventional KT function $G_z^{\rm KT}(t;\Delta_{\rm KT},\nu_{\rm KT})$ that serves as an approximation of $G_z^{\rm EA}(t)$. The proposed formulae successfully reproduce the anomalous temperature dependences of the linewidth $\Delta_{\rm KT}(T)$ and the fluctuation rate $\nu_{\text{KT}}(T)$ which are artifacts observed upon analyzing $\mu$SR data on ion diffusion using $G_z^{\rm KT}(t)$.  Applying this model to $\Delta_{\rm KT}$ and $\nu_{\rm KT}$ published in the literature of battery materials yields activation energy $E_{\rm i}$ and pre-exponential factor $\nu_{\rm i0}$ in the Arrhenius equation for ion diffusion, $\nu_{\rm i}(T)=\nu_{\rm i0} e^{-E_{\rm i}/kT}$, that are in excellent agreement with those deduced using other techniques, transforming peculiar artifacts into a powerful diagnostic tool.
\end{abstract}

\maketitle

Elucidating the microscopic mechanism of ion diffusion in solids is important for the development of next-generation energy storage materials including batteries. Muon spin rotation and relaxation ($\mu$SR) under zero/longitudinal external field (ZF/LF) has been used  as a local probe for such ion dynamics because of its high sensitivity to local magnetic fields exerted from nuclear magnetic moments of jumping ions.
Conventionally, the observed $\mu$SR time spectra have been analyzed by the least-square ($\chi^2$) curve fits using the dynamical Gaussian Kubo-Toyabe (KT) relaxation function $G_z^{\text{KT}}(t; \Delta_{\text{KT}},\nu_{\text{KT}})$ \cite{Hayano:79}, where the derived fluctuation rate $\nu_{\text{KT}}$ has been interpreted as the jump rate of the host ions ($\nu_{\rm i}$) that induce the fluctuation of the Gaussian linewidth $\Delta_{\text{KT}}$ in its full amplitude \cite{Sugiyama:09,Sugiyama:10,Sugiyama:11,Baker:11,Sugiyama:12,Mansson:13,Ashton:14, Mansson:14,Sugiyama:15,Amores:16,Umegaki:17,Laveda:18,Sugiyama:20,Benedek:20,McClelland:20,Ohishi:22a,Ohishi:22b,Umegaki:22,Ohishi:23,Forslund:25}.

However, the recent insight that a single ion-jump event modulates only a portion of the internal magnetic fields acting on muons from a large number of ions, and that the lineshape of spin relaxation in such case should exhibit qualitatively different behavior from that of the standard dynamical KT function has led to the development of an ``extended'' Kubo-Toyabe relaxation function, $G_z^{\text{EA}}(t)$, based on the Edwards-Anderson (EA)-type autocorrelation for the fluctuating local field $B(t)$, 
\begin{equation}
\gamma_\mu^2\langle B(t)B(0)\rangle =\Delta^2[(1-Q) + Q e^{-\nu_i t}]e^{-\nu_\mu t},\label{EA}
\end{equation} 
where $\langle ...\rangle$ means the statistical average over the canonical ensemble, $\gamma_\mu$ is the muon gyromagnetic ratio ($=2\pi\times135.53$ MHz/T), $\Delta$ is the linewidth determined by the density distribution $n(B)$ in the quasistatic limit, $Q$ ($<1$) is the fraction of the fluctuating local fields originating from the host ion jumps, and $\nu_{\rm i}$ ($\nu_\mu$) is the jump rate of ions (muons) \cite{Ito:24,Ito:26}. Note that $1-Q$ is not attributed solely to stationary ions, but also includes contributions from Li ions that did not jump during the single event of fluctuation in $B(t)$; no distinction is made between these two \cite{Ito:26}.  Crucially, when host ions migrate, only part of the linewidth $\sqrt{Q}\Delta$ fluctuates with ion motion, whereas the full linewidth $\Delta$ fluctuates for the muon motion irrespective of $\nu_{\rm i}$. This makes it feasible to experimentally distinguish the ion diffusion from muon self-diffusion. 

It has been shown in our recent simulations that, while the standard KT function serves as a reasonable approximation of $G_z^{\text{EA}}(t;\Delta;\nu_{\rm i},\nu_\mu,Q)$ with $Q < 1$, the parameters of the KT function exhibit peculiar behavior in terms of $\Delta$ and $\nu_{\rm i}$: $\nu_{\text{KT}}$ displays a conspicuous peak at a certain ion jump rate $\nu_{\rm i}^*$ which is accompanied by a significant reduction in $\Delta_{\text{KT}}$ for $\nu_{\rm i}\gtrsim\nu_{\rm i}^*$, indicating highly non-trivial relationship between two sets of parameters, $\Delta$, $\nu_{\rm i}$ and $\Delta_{\text{KT}}$, $\nu_{\text{KT}}$ \cite{Kadono:26}. Moreover, provided that $\nu_{\rm i}$ monotonically increases with temperature $T$ (as is the case of ion diffusion that obeys the Arrhenius equation), this immediately leads to the prediction that $\nu_{\text{KT}}$ should exhibit a similar peak at a temperature $T^*$ where $\nu_{\rm i}(T^*)=\nu_{\rm i}^*$, which is accompanied by a significant reduction in $\Delta_{\text{KT}}$ for $T\gtrsim T^*$. It is noteworthy that such behavior has been indeed confirmed in the bulk of the earlier literature on battery materials \cite{Kadono:26}. 

Nonetheless, there remains a major challenge of reanalyzing vast amounts of historical data using $G_z^{\text{EA}}(t)$ in order to derive the proper ion jump frequency. In this Letter, we overcome this challenge by introducing a simple empirical model that analytically relates the apparent profiles of $\Delta_{\rm KT}(T)$ and $\nu_{\rm KT}(T)$ to those of $\Delta(T)$,  $\nu_i(T)$, and $\nu_\mu(T)$ with $T$ as an internal parameter. 

The anomalous behavior of the KT function parameters in mimicking $G_z^{\rm EA}(t)$ stems from the $Q$ component that induces a maximum in the spin relaxation rate at $\nu_{\rm i}^*$ and subsequent reduction (motional narrowing) for $\nu_i>\nu_i^*$.  $G_z^{\rm KT}(t)$ reproduces the corresponding $\mu$SR lineshape by reducing $\Delta_{\rm KT}$ and $\nu_{\rm KT}$ as effective parameters. Since $\Delta_{\rm KT}$ is independent of $\nu_\mu$ by definition, the decay in the contribution of the $Q$ component to $\Delta_{\rm KT}$ can be approximated in the spirit embodied for the general theory of spin relaxation in NMR \cite{Bloembergen:48,Kubo:54}:
\begin{equation}
\Delta_{\rm KT}^2\approx\Delta^2\left[\frac{2(e^{-\nu_{\rm i}}-1+\nu_{\rm i})}{\nu_{\rm i}^2}Q+(1-Q)\right],\label{tsm}
\end{equation}
where the first term is obtained by assuming that the effective linewidth is determined by the spin relaxation rate ($\propto1/T_1$) \cite{Kubo:54}. 
Regarding $\nu_{\rm KT}$,  the anomalous peak is approximated as
\begin{equation}
\nu_{\text{KT}} \approx  \Gamma(\nu_{\rm i}) +\nu_\mu \equiv \frac{2\Delta_{\rm p}^2 \nu_{\rm i}}{2\Delta_{\rm p}^2 + \nu_{\rm i}^2} +\nu_\mu ,\label{nuKT}
\end{equation}
where $\Delta_{\rm p}$ is a phenomenological parameter proportional to $\sqrt{Q}\Delta$ and governs the peak position determined by the condition $\nu_{\rm i}=\nu_{\rm i}^*=\sqrt{2}\Delta_{\rm p}$.  Note that the first term in Eq.~(\ref{nuKT}) is based on the formula for $1/T_1$ ($\propto\nu/(\omega^2+\nu^2)$ \cite{Bloembergen:48}, where the Zeeman term $\omega^2$ is substituted by $2\Delta_{\rm p}^2$ in the zero field limit): it satisfies the conditions that $\Gamma(\nu_{\rm i})\approx\nu_{\rm i}$ for $\nu_{\rm i}\ll\nu_{\rm i}^*$ and $\Gamma(\nu_{\rm i})\approx2\Delta_{\rm p}^2/\nu_{\rm i}$ for $\nu_{\rm i}\gg\nu_{\rm i}^*$ (i.e., in the motional narrowing limit).  Assuming that both host ion and muon diffusion are thermally activated processes, their jump rates follow the Arrhenius equations,
\begin{equation}
\nu_\alpha(T) = \nu_{\alpha0} \exp\left(-\frac{E_\alpha}{k_{\text{B}}T}\right),\:\:(\alpha=\mu,{\rm i}) \label{nui}\\
\end{equation}
where $E_\mu$ and $E_{\rm i}$ are the activation energies for the ion and muon jump motions, respectively, and $\nu_{\rm i0}$ and $\nu_{\mu0}$ are the respective pre-exponential factors. 

 \begin{figure}[t]
\centering
\includegraphics[width=0.9\linewidth]{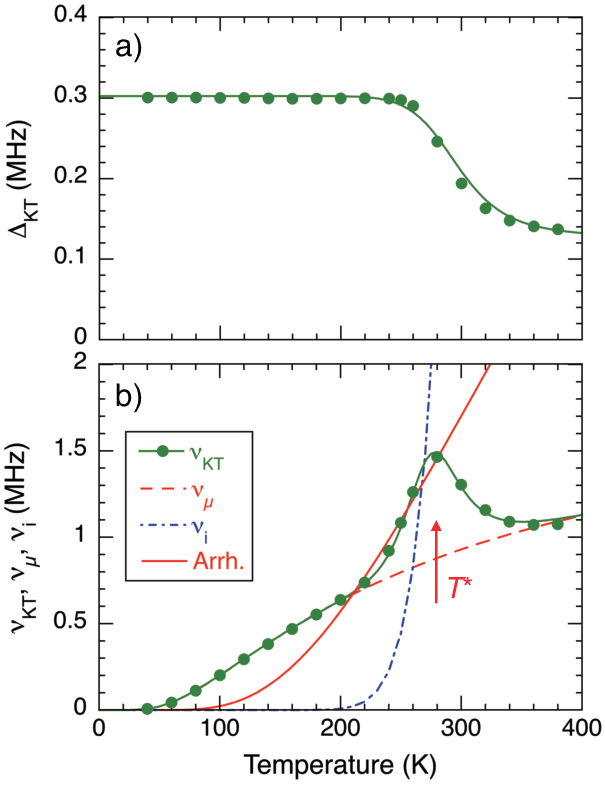}
\caption{(a) Linewidth and (b) fluctuation frequency versus temperature obtained by simulations (solid circles: deduced by curve fits using the standard KT function applied to the time spectra generated by $G_z^{\rm EA}(t)$ with $\nu_{\rm i}$ and $\nu_\mu$ given by Eq.~(\ref{nui})). Solid line in (a) is the best fits of $\Delta_{\rm KT}$ with Eq.~(\ref{tsm}). Solid lines in (b) are best fits of $\nu_{\rm KT}$ with Eq.~(\ref{nuKT}) and with Eq.~(\ref{nui}) for 200--280 K (``Arrh.''), respectively.  $\nu_{\rm \mu}$ (dashed line) and $\nu_{\rm i}$ (dot-dashed line) in (b) are drawn with parameters deduced by the fit (see Table \ref{tab1}).   }
\label{dkntm}
\end{figure}

\begin{table}[b]
\begin{tabular}{ccccccc}
\hline\hline
  & $\Delta_{\rm p}$ & $\nu_{\mu0}$ & $E_\mu$ & $\nu_{\rm i0}$  & $E_{\rm i}$ & $Q$ \\
  & (MHz) & (MHz) &  (eV) & (s$^{-1}$) &  (eV) & \\
\hline
Sim. & -- & 2.0 & 0.02 & $5.0\times10^{12}$ &  0.35 & 0.8 \\ 
\hline
$\Delta_{\rm KT}$$^a$  & -- & -- & -- & $1(1)\times10^{13}$ & 0.35(2) & 0.82(2) \\
$\nu_{\rm KT}$$^b$ & 0.89(2) & 1.97(5) & 0.020(1) & $2(1)\times10^{12}$ & 0.34(1) & -- \\
\hline
$\nu_{\rm KT}$$^c$  & -- & -- & -- & $1.5(3)\times10^{7}$ & 0.057(5) & -- \\
\hline\hline
\end{tabular}
\caption{Comparison of parameters obtained from curve fits of $\Delta_{\rm KT}$ and $\nu_{\rm KT}$ versus temperature with those assumed in the numerical simulations of $\mu$SR time spectra (Sim.). $a$ with Eq.~(\ref{tsm}); $b$ with Eq.~(\ref{nuKT}); $c$ with Eq.~(\ref{nuKTi}).}\label{tab1}
\end{table}

To validate this approach, we performed numerical benchmarks, where $\mu$SR time spectra (ZF and LF = 1 mT) were generated by $G_z^{\text{EA}}(t;\Delta;\nu_{\rm i},\nu_\mu,Q)$ (with $\Delta=0.3$ MHz, $Q =0.8$) and analyzed by $\chi^2$ curve fits using the standard KT function to deduce $\Delta_{\rm KT}$ and $\nu_{\text{KT}}$ \cite{Kadono:26}; the authenticity of input parameters for Eq.~(\ref{nui}) (shown in Table \ref{tab1}) is discussed in detail elsewhere \cite{Ito:26}. Figure \ref{dkntm}(a) shows $\Delta_{\text{KT}}$ versus temperature and result of curve fit using Eqs.~(\ref{tsm}) and (\ref{nui}), which yields $Q=0.82(2)$, $\nu_{\rm i0}\approx1(1)\times10^{13}$ s$^{-1}$, and $E_{\rm i}=0.35(2)$ eV.  The result of fits for $\nu_{\text{KT}}$ using Eqs.~(\ref{nuKT}) and (\ref{nui}) is also shown in Fig.~\ref{dkntm}(b), where our simple empirical model reproduces the complex behavior of $\nu_{\text{KT}}(T)$ remarkably well. The fit yields $\nu_{\rm i0} = 2(1)\times10^{12}$~s$^{-1}$ and $E_{\rm i} = 0.34(1)$~eV, which perfectly reproduce the original values within errors (the same is true for $\nu_{\mu0}$ and $E_\mu$). Furthermore, the phenomenological parameter $\Delta_{\rm p} = 0.89(2)$~MHz is consistent with that evaluated via the simulation of $\Delta_{\rm p} \simeq 3\sqrt{Q}\Delta=0.82$ MHz \cite{Kadono:26}, confirming that Eq.~(\ref{nuKT}) captures the underlying mechanism of the spin relaxation governed by the EA-type autocorrelation with $Q<1$. (The validity of this approach was also confirmed for the simulation data with $Q=0.5$.) 

In contrast, when assuming $\nu_{\mu}=0$ and 
\begin{equation}
\nu_{\rm KT}=\nu_{\rm i} = \nu_{\rm i0} \exp\left(-\frac{E_{\rm i}}{k_{\text{B}}T}\right)\label{nuKTi},
\end{equation}
fitting the data within the limited temperature range of 200--280 K (see ``Arrh.'' in Fig.~\ref{dkntm}(b)) yields $\nu_{\rm i0} = 1.5(3)\times10^7$ s$^{-1}$ and $E_{\rm i} = 0.057(5)$ eV which are significantly underestimated.  In the following, we demonstrate the usefulness of these formulae by applying them to experimental datasets available in the literature for typical battery materials. 

In order to verify the validity of the empirical model, we selected three examples, C$_6$Li, Li$_x$CoO$_2$ and Li$_x$FePO$_4$, for the reasons described below. The merit of C$_6$Li is that there are a variety of experimental results available for crosschecking the $\mu$SR result \cite{Umegaki:17}, including quasielastic neutron scattering and Li-NMR \cite{Langer:13}. Li$_x$CoO$_2$ is the first material for which Li-ion diffusion via $\mu$SR was discussed \cite{Sugiyama:11}, and research using operando-$\mu$SR measurements has continued since then \cite{Ohishi:22b}. Li$_x$FePO$_4$ is also a material of great interest, and similar results have been reported in multiple $\mu$SR studies \cite{Sugiyama:11,Baker:11,Sugiyama:12,Ashton:14}.

\begin{figure}[t]
\centering
\includegraphics[width=0.95\linewidth]{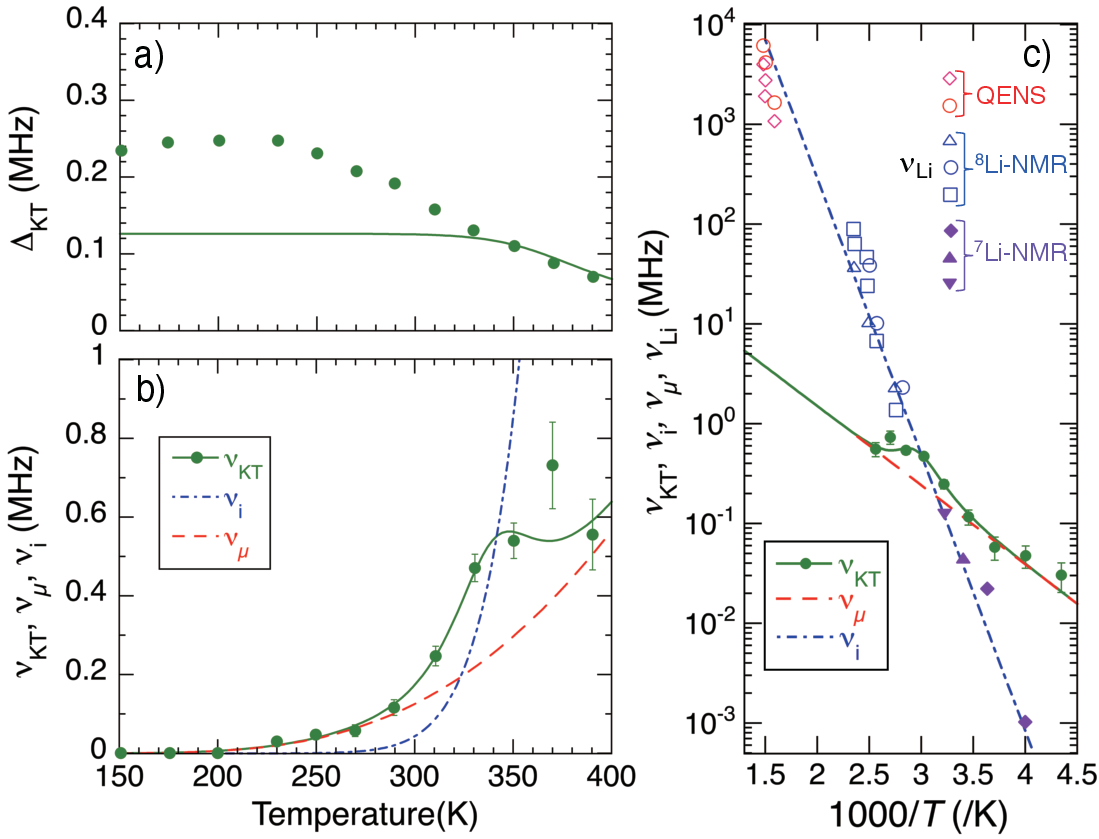}
\caption{(a) Linewidth $\Delta_{\rm KT}$ and (b) fluctuation frequency $\nu_{\rm KT}$ in C$_6$Li (solid circles: after Ref.[\onlinecite{Umegaki:17}]). Solid line in (a) shows the behavior of $\Delta_{\rm KT}$ associated with $\nu_{\rm i}$ predicted from the $\nu_{\rm KT}$ fit, and that in (b) is the best fit of $\nu_{\rm KT}$ using Eq.~(\ref{nuKT}), where $\nu_\mu$ (dashed line) and $\nu_{\rm i}$ (dot-dashed line, attributed to Li ion jump) are drawn with parameters deduced by the fit (see Table \ref{tab2}).  (c) Comparison of $\nu_{\rm i}$ and $\nu_\mu$ with Li jump frequency $\nu_{\rm Li}$ observed by other techniques (after Ref.[\onlinecite{Langer:13}]).  }
\label{C6Li}
\end{figure}

As a first example, let us look into the results of the $\mu$SR experiment in lithium-intercalated graphite C$_6$Li \cite{Umegaki:17}.  As shown in Fig.~\ref{C6Li}(b), $\nu_{\text{KT}}$ shows a broad peak (or a shoulder) at $T^*\approx350$ K, while other experimental techniques have coherently suggested a monotonically increasing trend in $\nu_{\rm i}$.  The curve fits of $\nu_{\rm KT}(T)$ using Eq.~(\ref{nui}) was adopted over a specific temperature region below $T^*$ (270--330 K) to yield $E_{\rm i} =0.270(5)$ eV \cite{Umegaki:17}, which was significantly smaller than 0.55 eV inferred from $^7\text{Li}$-NMR \cite{Langer:13}. The corresponding $\nu_{\rm i0}$ was in the range of 10$^8$--10$^9$ s$^{-1}$, which was also smaller than the reported value by orders of magnitude.

As shown in Fig.~\ref{C6Li}(b), we have been successfully reproduced $\nu_{\rm KT}(T)$ including the broad peak around $T^*\approx350$ K by curve fit using Eqs.~(\ref{nuKT})--(\ref{nui}) and to deduce the parameters for both $\nu_\mu$ and $\nu_{\rm i}$: the results are summarized in Table \ref{tab2}. The obtained activation energy $E_{\rm i} = 0.55(19)$~eV matches that of Li ion diffusion determined by $^7\text{Li}$-NMR \cite{Langer:13} with reasonable accuracy, and $\nu_{\rm i0}$ is also found to be in the range of $10^{13}$--$10^{14}$~s$^{-1}$, aligning perfectly with both the $^7$Li-NMR result as well as typical Debye frequency of the host lattice ($\Theta_D=300(30)$ K \cite{Delhaes:76}, $k_B\Theta_D/h\approx6.3(6)\times10^{12}$ s$^{-1}$).  Regarding muon motion, the relatively small $\nu_{\mu0}$ ($=10^7$--10$^8$ s$^{-1}$) and $E_\mu$ ($\approx$0.16 eV) suggest that the jump process is dominated by the phonon-assisted tunneling \cite{Fukai:05}.

\begin{table}[b]
\begin{tabular}{cccccc}
\hline\hline
  & $\Delta_{\rm p}$ (MHz) & $\nu_{\mu0}$ (MHz) & $E_\mu$ (eV) & $\nu_{\rm i0}$ (s$^{-1}$) & $E_{\rm i}$ (eV) \\
\hline
 This work & 0.39(8) & 52(31) & 0.157(15) & $7\times10^{13}$ & 0.55(19)  \\
Li-NMR & -- & -- & -- & $5\times10^{14}$ & 0.55  \\
Ref.[\onlinecite{Umegaki:17}] & -- & -- & -- & $5\times10^{9}$ & 0.270(5)  \\
\hline\hline
\end{tabular}
\caption{Fitting parameters for muon and Li-ion jump motion obtained from curve fits of $\nu_{\rm KT}$ versus temperature using Eq.~(\ref{nuKT}). $\nu_{\rm i0}$ and $E_{\rm i}$ reported in Li-NMR literature \cite{Langer:13} are quoted for comparison.}\label{tab2}
\end{table}

Meanwhile, $\Delta_{\rm KT}(T)$ shown in Fig.~\ref{C6Li}(a) decreases with $T$ from around 220 K which is considerably lower than $T^*$, seemingly in contradiction with the empirical model. However, this apparent inconsistency is resolved by considering the magnitude of $\Delta_{\rm p}$ obtained from the $\nu_{\rm KT}$ analysis (see Table \ref{tab2}). Since the linewidth subjected to the fluctuation due to Li-ion jumps is estimated to be $\sqrt{Q}\Delta\approx \Delta_{\rm p}/3=0.13$ MHz, only a part of $\Delta_{\rm KT}$ corresponds to that described by Eq.~(\ref{tsm}), as indicated by the solid line in Fig.~\ref{C6Li}(a) (where $Q=0.9$ and $\sqrt{Q}\Delta=0.126$ MHz obtained by fit with $\nu_{\rm i0}$ and $E_{\rm i}$ fixed to those in Table \ref{tab2}). Furthermore, considering that Li ions are virtually stationary below $\approx$300 K, it is inferred that the decrease in $\Delta_{\rm KT}$  at 220--350 K is primarily due to muon diffusion (see below). 

 \begin{figure}[t]
\centering
\includegraphics[width=1.0\linewidth]{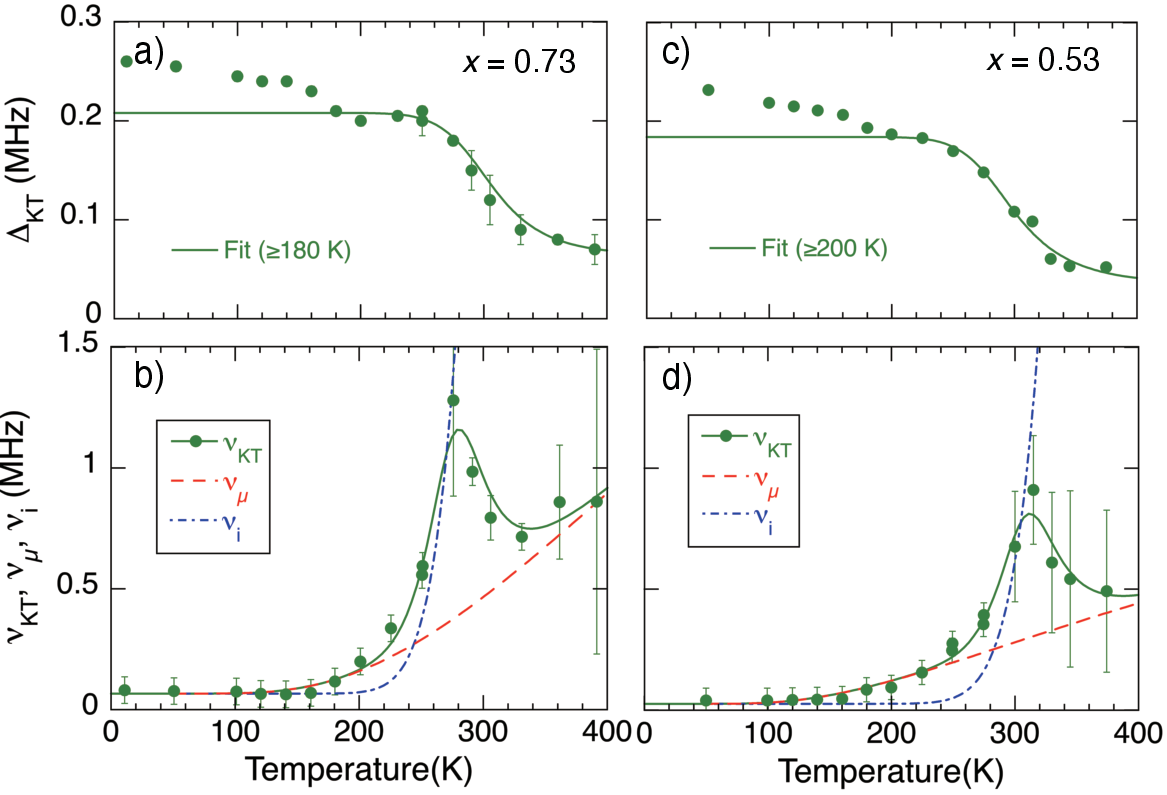}
\caption{(a) (c) Linewidth $\Delta_{\rm KT}$ and (b) (d) fluctuation frequency $\nu_{\rm KT}$ versus $T$ observed in Li$_x$CoO$_2$ (solid circles: after Ref.[\onlinecite{Sugiyama:09}]). Solid line in (a) (c) shows the fit results using Eq.~(\ref{tsm}) for the data at higher temperatures. Solid line in (b) (d) is the best fit of $\nu_{\rm KT}$ using Eq.(\ref{nuKT}), where $\nu_{\rm \mu}$ (dashed line) and $\nu_{\rm i}$ (dot-dashed line) are drawn with parameters deduced by the fit (see Table \ref{tab3}). }
\label{LiCO}
\end{figure}

\begin{table}[b]
\begin{tabular}{c|c|ccccc}
\hline\hline
\multirow{3}{*}{$\Delta_{\rm KT}$}& $x$  & $\Delta$ (MHz) & $Q$ & $\nu_{\rm i0}$ (s$^{-1}$) & $E_{\rm i}$ (eV) \\
\cline{2-6}
&0.73& 0.208(3) & $0.91(5)$ & 2(6)$\times10^{13}$ & 0.36(9)  \\
&0.53 & 0.184(4) & 0.96(2)& $1(3)\times10^{13}$ & 0.34(5)  \\
\hline\hline
\end{tabular}

\begin{tabular}{c|c|ccccc}
\hline\hline
\multirow{3}{*}{$\nu_{\rm KT}$} &$x$ & $\Delta_{\rm p}$ (MHz) & $\nu_{\mu0}$ (MHz) & $E_\mu$ (eV) & $\nu_{\rm i0}$ (s$^{-1}$) & $E_{\rm i}$ (eV) \\
\cline{2-7}
&0.73& 1.09(12)& 7.4(4.8) & 0.076(22) & $2.3\times10^{12}$ & 0.34(6)  \\
&0.53 & 0.72(7) & 1.8(1.0) & 0.051(15) & $4.6\times10^{12}$ & 0.41(8) \\
\hline
\multirow{2}{*}{Ref.\cite{Sugiyama:09}}& 0.73& --& -- & -- & $4.2\times10^{7}$ & 0.093  \\
&0.53 & -- & -- & -- & $2.3\times10^{7}$ & 0.096 \\
\hline\hline
\end{tabular}
\caption{Fitting parameters for muon and Li-ion jump motion in Li$_x$CoO$_2$ with $x=0.53$ and 0.73 obtained from curve fits of $\Delta_{\rm KT}$ and $\nu_{\rm KT}$ versus temperature using Eqs.~(\ref{tsm})--(\ref{nui}), respectively. }\label{tab3}
\end{table}

The second example is $\text{Li}_{x}\text{CoO}_2$, a well-known cathode material, for which results of the analysis using the standard KT function on the samples with $x=0.73$ and $0.53$  \cite{Sugiyama:09} are quoted in Fig.~\ref{LiCO}. Interestingly, $\nu_{\rm KT}(T)$ shows a pronounced peak near $T^*\approx270$ K for $x=0.73$  and $\approx$300 K for $x=0.53$, respectively, and $\Delta_{\rm KT}(T)$ decreases significantly at temperatures higher than $T^*$ in both cases. However, these characteristics had drawn little attention until very recently \cite{Kadono:26}; instead, the temperature dependence of $\nu_{\rm KT}$ below $T^*$ was attributed to Li-ion diffusion (with the muon assumed to be immobile). 

Now, as is found in Figs.~\ref{LiCO}(b) and \ref{LiCO}(d), the curve fits of $\nu_{\text{KT}}$ using Eq.~(\ref{nuKT}) show excellent agreement with data, yielding parameter values shown in Table \ref{tab3}. Moreover, these results indicate that the decrease in $\Delta_{\rm KT}$ for 150--200 K is not related with Li ion motion. In fact, curve fits of $\Delta_{\rm KT}$ using Eq.~(\ref{tsm}) for the relevant temperature range yield $\nu_{\rm i}$ and $E_{\rm i}$ which are in good agreement with those deduced from $\nu_{\rm KT}$, strongly highlighting the self-consistency and reliability of our empirical model. These results are also in line with Li-NMR which suggests $E_{\rm i}\approx0.3$ eV for $x=1.0$ and 0.8 \cite{Nakamura:06}. The pre-exponential factors are also consistent with that estimated from the reported $\Theta_D$ ($\approx300$ K) \cite{Takahata:02}.  These results are in marked contrast with the previous analysis results using Eq.~(\ref{nui}) that yielded significantly smaller $\nu_{\rm i0}$ and $E_{\rm i}$ (see Table \ref{tab3}) \cite{Sugiyama:09}.

\begin{figure}[t]
\centering
\includegraphics[width=0.65\linewidth]{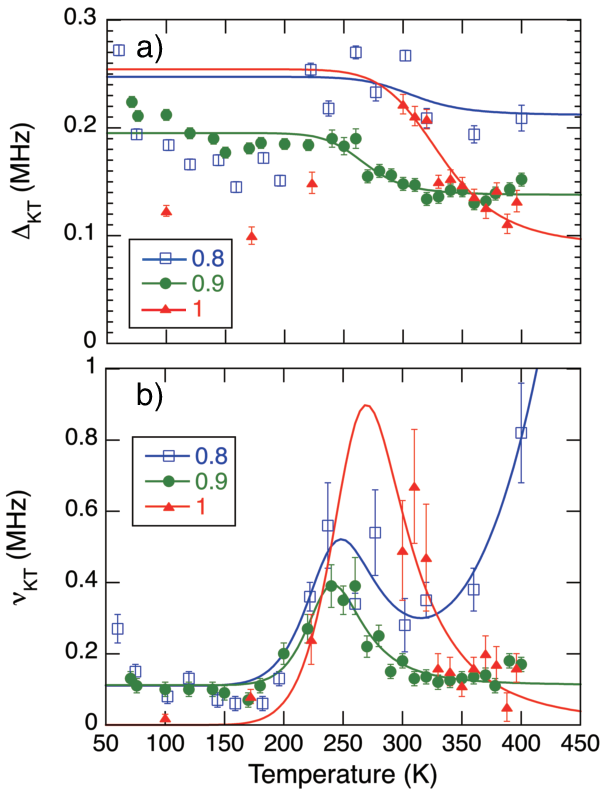}
\caption{(a) Linewidth $\Delta_{\rm KT}$ and (b) fluctuation frequency $\nu_{\rm KT}$ versus $T$ observed in Li$_x$FePO$_4$ for $x=0.8$ (squares), 0.9 (circles), and 1 (triangles) (after Ref.[\onlinecite{Baker:11}]).    Solid lines in (a) show fit results using Eq.~(\ref{tsm}), where the data for $T>200$ K and $>270$ K were used for $x=0.8$ and 1, respectively. Solid lines in (b) are the best fits of $\nu_{\rm KT}$ using Eqs.~(\ref{nuKT})--(\ref{nui}), where a $T$-independent offset $\nu_{\rm c}$ was considered for $x=0.8$ and 0.9. (see Table \ref{tab4} for the parameter values.)  }
\label{LFP}
\end{figure}

The last example is the olivine-type Li$_x$FePO$_4$, where the analysis results using the standard KT function reveal a more complex situation. Figure \ref{LFP} shows plots reproducing $\Delta_{\rm KT}(T)$ and $\nu_{\rm KT}(T)$ reported for $x = 0.8$, 0.9, and 1 \cite{Baker:11}. Of particular interest is that $\nu_{\rm KT}$ exhibits a distinct peak near $T^*\approx250$--300 K in all samples, where the decrease in $\nu_{\rm KT}$ for $T>T^*$ was attributed to a ``decrease in sensitivity'' in the literature, and $\nu_{\rm KT}$ for $T<T^*$ was analyzed using Eq.~(\ref{nui}) to yield $E_{\rm i} \approx 0.10$ eV. Similar values are reported in other literature \cite{Sugiyama:11,Sugiyama:12}, which are 1.5--7 times smaller than those evaluated using various other techniques and first-principles density functional theory calculations \cite{Morgan:04,Islam:05}. However, the maximum value of $\nu_{\rm KT}$ in the $x = 0.9$ sample, for example, is $\sim$0.4 MHz which is only about twice of $\Delta_{\rm KT}$, and as long as the $\mu$SR spectra follow the standard KT function, $\nu_{\rm KT}$ remains in a region where $\nu_{\rm KT}\approx\nu_{\rm i}$. 

The results of analyzing $\nu_\text{KT}(T)$ versus $T$ using Eqs.~(\ref{tsm})--(\ref{nui}) are shown as solid lines in Fig.~\ref{LFP} with derived parameter values in Table \ref{tab4}. Here, a temperature-independent offset $\nu_{\rm c}$ was considered in Eq.~(\ref{nuKT}) to reproduce the data for $x=0.8$ and 0.9 (see below).  A remarkable feature compared to the two preceding examples is that $E_\mu$ is larger than $E_{\rm i}$ and virtually infinite in $x=1$ as $\nu_\mu\approx0$. Although the pre-exponential factors, $\nu_{\rm i0}$ for the samples with $x=0.8$ and 0.9 are larger than $\nu_{\mu0}$, the difference is one order of magnitude at most. This would have made it difficult to distinguish between muon and ion diffusion based solely on the magnitudes of these parameters. However, the peak structures in $\nu_\text{KT}(T)$, which is not expected for muon diffusion (as $Q=0$), provides strong evidence for the Li diffusion.

\begin{table}[t]
\begin{tabular}{c|c|cccc}
\hline\hline
\multirow{2}{*}{$\Delta_{\rm KT}$}&$x$  & $\Delta$ (MHz) & $Q$ & $\nu_{\rm i0}$ (s$^{-1}$) & $E_{\rm i}$ (eV) \\
\cline{2-6}
&0.8& 0.247(24) & 0.27(3) & $3(1)\times10^{12}$ & 0.34(3) \\
&0.9& 0.195(2) &0.50(2) & $2(6)\times10^{12}$ & 0.32(9) \\
&1& 0.254(51) &0.87(14) & $2(8)\times10^{12}$ & 0.34(21) \\
\hline\hline
\end{tabular}

\begin{tabular}{c|c|ccccc}
\hline\hline
\multirow{4}{*}{$\nu_{\rm KT}$} &$x$ & $\Delta_{\rm p}$ (MHz) & $\nu_{\mu0}$ (MHz) & $E_\mu$ (eV) & $\nu_{\rm i0}$ (s$^{-1}$) & $E_{\rm i}$ (eV) \\
\cline{2-7}
&0.8& 0.57(13)& $2\times10^3$ & 0.27(13) & $10^{10}$--$10^{11}$ & 0.19(9)  \\
&0.9 & 0.42(2) & $1.4\times10^4$ & 0.4(2) & $10^{11}$--$10^{12}$ & 0.23(2) \\
&1 & 1.3(5) & -- & -- & $10^{11}$--$10^{12}$ & 0.23(4) \\
\hline
\multirow{3}{*}{Ref.\cite{Baker:11}} &0.8& --& -- & -- & $3\times10^{8}$ & 0.13  \\
&0.9 &-- & -- & -- & $2\times10^{7}$ & 0.08 \\
&1 & -- & -- & -- & $5\times10^{6}$ & 0.06 \\
\hline\hline
\end{tabular}
\caption{Fitting parameters for muon and Li-ion jump motion Li$_x$FePO$_4$ with $x=0.8$, 0.9, and 1 obtained from curve fits of $\nu_{\rm KT}$ and $\Delta_{\rm KT}$ versus temperature using Eqs.~(\ref{tsm})--(\ref{nui}), respectively. The offset $\nu_{\rm c}$ was $0.11(4)$ for $x=0.8$ and $0.11(1)$ for 0.9.}\label{tab4}
\end{table}

Coming to the linewidth $\Delta_{\rm KT}(T)$ shown in Fig.~\ref{LFP}(a), that of the $x = 0.9$ sample decreases for $T>T^*\approx240$ K where the $\nu_{\rm KT}$ exhibits a peak, which is again perfectly in line with our model. As shown in Table \ref{tab4}, the parameters obtained by fitting this change using Eq.~(\ref{tsm}) are consistent with those derived from the analysis of $\nu_{\rm KT}(T)$ versus $T$. However,  the observed decrease for $T<T^*$ with decreasing $T$ in the samples with $x = 0.8$ and $x = 1$ differs significantly from previous examples (in fact, a slight tendency toward the same behavior is also observed for $x = 0.9$), whose origin is yet to be clarified. 

The remaining issue here is the origin of the $T$ dependence of $\Delta_{\rm KT}$ at $T<T^*$, which is observed in all three cases.  It has been shown that such changes in linewidth can be understood as trapping and/or detrapping of muons at sites with different $\Delta_{\rm KT}$ values in samples containing impurities and/or defects \cite{Borghini:78,Kehr:78,Kehr:82}; Li vacancies present in high concentrations in these battery material are considered to be a possible cause. Specifically, for C$_6$Li and Li$_x$CoO2, concrete examples of $\Delta_{\rm KT}$ analysis that take this point into account are presented in Supplemental Material (SM) \cite{SM}.

Regarding the origin of $\nu_{\rm c}$, one possibility is that the deviation in the actual spin relaxation from the KT function may be compensated by an additional variation in $\nu_{\rm KT}$ in the curve fit. A slight contribution of $\nu_{\rm c}$ is also observed for Li$_x$CoO$_2$. On the other hand, while theoretical calculations predicted the possibility of such a discrepancy for C$_6$Li \cite{Pratt:26}, it does not appear to be observed in Fig.~\ref{C6Li}(b).

In addition to the three cases described above, we conducted similar analyses on several sets of data from previous literature including FAPbI$_3$ (hybrid perovskite), LaH$_{2.5}$O$_{0.25}$, and Li$_y$Nb$_{0.1}$Ti$_{0.9}$ ($y=0,1$), and obtained generally satisfactory results (see SM \cite{SM}). Based on these findings, it is now concluded that the validity of our model has been sufficiently demonstrated. 

The significance of the present model lies in its ability to serve as a universal inverse-problem solution procedure, not only for retrieving information on ion diffusion from existing $\mu$SR literature but also  for the future $\mu$SR studies on the relevant fields where the ion dynamics is a focus of interest. Instead of undertaking demanding analysis of time-differential $\mu$SR spectra using $G_z^{\text{EA}}(t)$ with multiple parameters, researchers can now extract the genuine ion-dynamics parameters (and muon characteristics) via routine curve fits using the standard KT relaxation function. This procedure effectively bypasses the mathematical complexity of spin relaxation functions while retaining the underlying physics of the Edwards-Anderson-type correlation.

Finally, it should be noted that the reanalysis of Li$_{0.73}$CoO$_2$ data using the recently proposed spin relaxation model that consider the correlated jump motion between Li ions and muons by introducing the retarded memory torque reproduces the reduction in linewidth for $T>T^*$ and the peak in the fluctuation frequency around $T\approx T^*$ \cite{Arguelles:26}, suggesting the irrelevance of the model to these anomalies.

In summary, we have established a highly practical empirical model with which the information on host ion dynamics can be deduced from the temperature dependence of fitting parameters for the standard Kubo-Toyabe relaxation function which is routinely used in the analysis of $\mu$SR spectra.  This approach resurrects the hidden physical value of extensive historical $\mu$SR datasets, providing an accessible and robust procedure for evaluating local ion dynamics.

The author thanks J. Sugiyama and  P. Baker for helpful discussion and allowing access to their published data for the secondary use. Thanks are also to T. U. Ito, M. Hiraishi, K. Ohishi, and I. Umegaki for helpful discussion. 
This work was partially supported by the MEXT Program: Data Creation and Utilization Type Material Research and Development Project (Grant No. JPMXP1122683430).

\let\doi\relax
%

\clearpage

\setcounter{figure}{0}
\setcounter{table}{0}
\setcounter{equation}{0}
\renewcommand{\thefigure}{S\arabic{figure}}
\renewcommand{\thetable}{S\arabic{table}}
\renewcommand{\theequation}{S\arabic{equation}}

\begin{center}
{\large Supplementary Material}\\
  \vspace{2mm}
  for\\
  \vspace{2mm}
  {\large An empirical model for deriving information on ion dynamics from $\mu$SR data via\\  the conventional Kubo-Toyabe function analysis}\\
    \vspace{5mm}
Ryosuke Kadono\\
Muon Science Laboratory, Institute of Materials Structure Science, High Energy Accelerator Research Organization (KEK), Tsukuba, Ibaraki 305-0801, Japan
\end{center}

\subsection{Temperature dependence of $\Delta_{\rm KT}$ due to muon diffusion in inhomogeneous systems}
In general, muon diffusion in solid crystals is dominated by phonon-assisted tunneling mechanisms at temperatures below the Debye temperature. In this case, the activation energy for muon jump is far smaller than the potential barrier energy for over-barrier jump of Li ions, and muons begin to diffuse at lower temperatures than cations. The results of our analysis using the empirical model shown in Figs.~2 and 3 demonstrate that this also holds true for C$_6$Li and Li$_x$CoO$_2$. Specifically, only $\nu_\mu$ exhibits a finite value at $T < 270$ K for C$_6$Li, at $T < 200$ K for Li$_{0.73}$CoO$_2$, and at $T < 240$ K for Li$_{0.53}$CoO$_2$, while the Li-ion jump frequency $\nu_{\rm i}$ is negligible. Furthermore, these temperature regions coincide with those where $\Delta_{\rm KT}(T)$ deviates from the predictions of the empirical model. Therefore, the change in $\Delta_{\rm KT}(T)$ at those points can be uniquely attributed to muon diffusion.

In the battery materials where the spatial Li ion distribution is inhomogeneous, the coordination of Li ions at muon sites varies from site to site. This suggests that the interstitial potential $V({\bm r})$ experienced by the muon ($\propto$ the formation energy for muon-Li-vacancy complexes) is also expected to differ, and that the linewidth $\Delta({\bm r})$ (which is primarily determined by the number of nearest neighboring (nn) ions with nuclear magnetic moments) also varies from site to site. 

\begin{figure*}[t]
\centering
\includegraphics[width=0.8\linewidth]{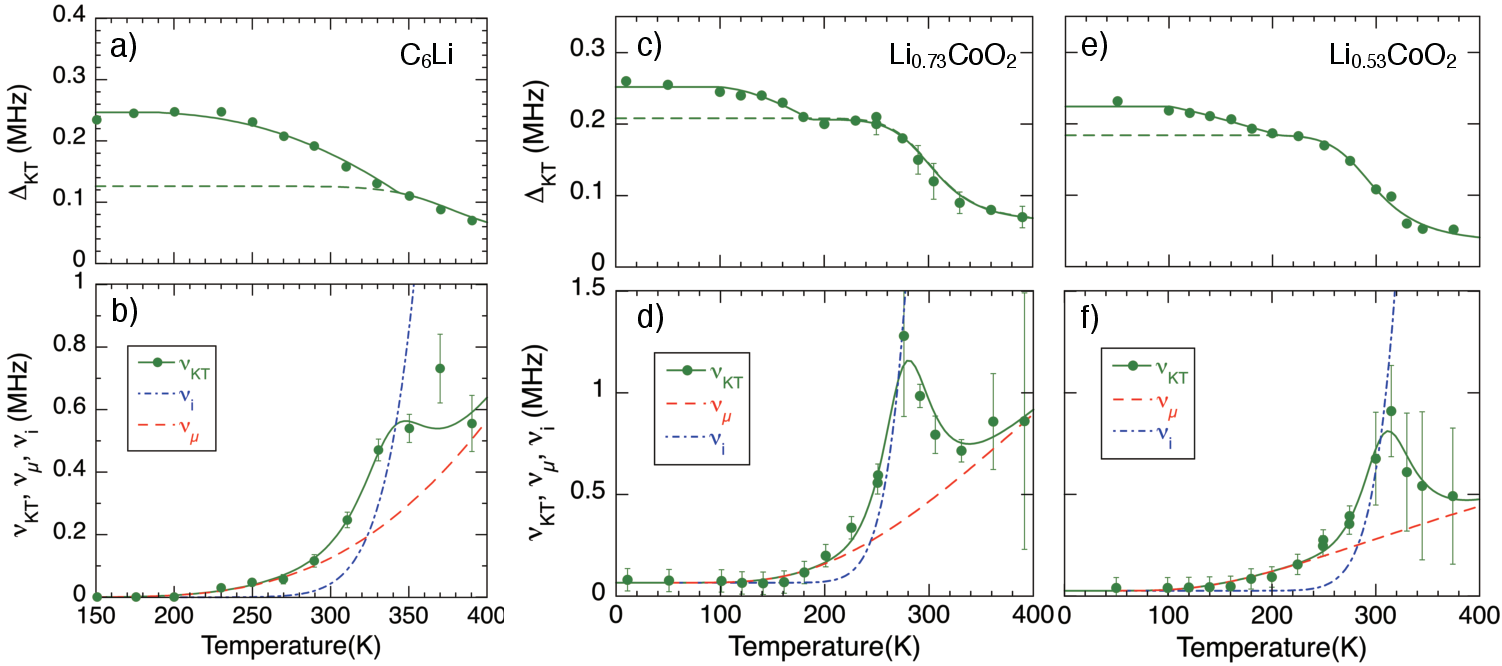}
\caption{Linewidth $\Delta_{\rm KT}$ and fluctuation frequency $\nu_{\rm KT}$ in C$_6$Li and Li$_x$CoO$_2$ quoted from the main text, where the solid (dashed) lines in (a), (c) and (e) are best fits after (before) implementing Eq.~(\ref{dltrap}). $\Delta_{\rm KT}(T)$ approaches $\Delta_{\rm av}$ (a spatial average of $\Delta({\bm r})$) for $T\rightarrow0$. ((b), (d), and (e) are unchanged.) } 
\label{dlc}
\end{figure*}

Now, muons immediately after implantation into these materials occupy interstitial sites randomly at low temperatures regardless of the spatial Li distribution, and $\Delta_{\rm KT}$ reflects the statistical average of $\Delta({\bm r})$ throughout the entire sample: $\langle\Delta({\bm r})\rangle_{\bm r}=\Delta_{\rm av}$. On the other hand, as the muon jump is thermally promoted, the occupation probability at sites with lower $V({\bm r})$ is likely to increase. (The migration of muons to sites with lower $V({\bm r})$ is expected to be completed in a short time at high temperatures, and jump diffusion between such low-$V$ sites is expected to become dominant.) The change in $\Delta_{\rm KT}$ induced by such an ``annealing process'' can be described by a diffusion-limited transition model, and the temperature dependence of the linewidth is expected to be given by the following equations:
\begin{equation}
\Delta_{\rm KT}^2(T)\approx(\Delta_0^2-\Delta^2)(1-x)e^{-x\tau_\mu\nu^*_\mu(T)}+\Delta^2,\label{dltrap}
\end{equation}
\begin{equation}
\nu^*_\mu(T)=\nu^*_{\mu0}e^{-E^*_\mu/k_BT}\label{nutrap}
\end{equation}
where $\Delta_0$ ($\Delta$) is the linewidth at sites with higher (lower) $V({\bm r})$, $x$ is the relative concentration of muon sites with $$\Delta_{\rm KT}(T\rightarrow0)=(1-x)\Delta_0^2+x\Delta^2=\Delta_{\rm av}^2,$$ $\tau_{\mu}$ is the average muon lifetime ($=2.2$ $\mu$s), and $\nu^*_{\mu}(T)$ is the muon jump frequency with $\nu^*_{\mu0}$ and $E^*_\mu$ being the pre-exponential factor and the activation energy. Here, we assume that $\Delta_0>\Delta$, which means that $V({\bm r})$ is lower for the sites with smaller number of nn Li ions. Using Eqs.~(\ref{dltrap}) and (\ref{nutrap}), we fitted $\Delta_{\rm KT}(T)$ with $\Delta_0$, $x$,  $\nu^*_{\mu0}$, and $E^*_\mu$ as free parameters, and the results are shown as solid lines in Fig.~\ref{dlc} and in Table \ref{tabdlc}. These fits reproduce the observed temperature dependence of $\Delta_{\rm KT}(T)$ excellently with reasonable parameter values;
$\nu^*_{\mu0}$ and $E^*_\mu$ are close to $\nu_{\mu0}$ and $E_\mu$ derived from the previous analysis on $\nu_{\rm KT}(T)$ for the respective samples, indicating that muon site migration driven by diffusion is a plausible scenario.

\begin{table}[h]
\begin{tabular}{ccccc}
\hline\hline
 & $\Delta_0$ (MHz) & $x$ & $\nu^*_{\mu0}$ (MHz) & $E^*_\mu$ (eV) \\
 \hline
 C$_6$Li & 0.35 & 0.41 & 251 & 0.15\\
 Li$_{0.73}$CoO$_2$ & 0.36 & 0.26 & 24 & 0.056\\
 Li$_{0.53}$CoO$_2$ & 0.31& 0.12 & 12 & 0.034\\
 \hline\hline
\end{tabular}
\caption{The parameters in Eqs.~(\ref{dltrap}) and (\ref{nutrap}) derived from curve fits of $\Delta_{\rm KT}(T)$, where $\Delta$ was fixed to those shown in Tables II and III.}\label{tabdlc}
\end{table}

Unlike the two examples described above, it is difficult to explain the temperature dependence of $\Delta_{\rm KT}(T)$ for $T<T^*$ in Li$_x$FePO$_4$ using such a simple scenario. However, the general principle that muons move more easily than cations via jump diffusion at low temperatures should still hold true here. Furthermore, considering that the complex temperature dependence observed in the linewidths of TF-$\mu$SR in solids containing impurities or defects is well understood in terms of muon diffusion-limited trapping and detrapping \cite{Borghini:78,Kehr:78,Kehr:82}, it is expected that a similar scenario of irreversible muon site change associated with the ``annealing process'' may explain the behavior of $\Delta_{\rm KT}$ below $T^*$ in Li$_x$FePO$_4$ as well.

\subsection{Rotational motion of cation molecules in FAPbI$_3$}

Hybrid organic-inorganic perovskites (HOIPs) are the subject of active research as promising solar cell materials. One reason believed to account for the high photoconversion efficiency of these materials is the long lifetime of photoexcited carriers. It is speculated that the rotational freedom of organic cation molecules plays a role in this phenomenon, and research using $\mu$SR has been conducted with a focus on this aspect \cite{Koda:22,Hiraishi:23}. 

Muons injected to HOIP come to rest to occupy specific site(s) within a unit cell, and exhibit spin relaxation as they sense random local magnetic fields $B(t)$ induced by the nuclear magnetic moments of the constituent nuclei on the perovskite lattice. This means that $B(t)$ consists of two components, one exerted from cation molecules which is fluctuating due to the local rotational motion within the cage, and the other from the PbI$_3$ lattice which is quasi-static. This strongly suggests that the time correlation of $B(t)$ is described by the Edwards-Anderson (EA) autocorrelation function. In fact, the $\mu$SR spectra observed in FAPbI$_3$ (FA = formamidinium, CH(NH$_2$)$_2$) in its quenched phase are the first examples with which the validity of the ``extended'' Kubo-Toyabe (KT) function $G_z^{\rm EA}(t)$ has been demonstrated \cite{Ito:24}. 

It has been revealed by this analysis that the sharp decrease in $\Delta_{\rm KT}$ upon analyzing the $\mu$SR spectra with the standard KT function \cite{Hiraishi:23}--which occurs as the temperature approaches the structural phase transition temperature of $\approx$140 K from low temperature--is due to the reduced contribution of the proton nuclear magnetic moments caused by the rapid rotation of the FA molecules (motional narrowing). However, the apparent peak in the fluctuation frequency $\nu_{\rm KT}$ observed in conjunction with the sharp decrease in $\Delta_{\rm KT}$ remained unexplained. Here, we show that the analysis of $\Delta_{\rm KT}$ and $\nu_{\rm KT}$ versus $T$ using our empirical model yields consistent results with those obtained directly from the $\mu$SR spectra using the $G_z^{\rm EA}(t)$ function.

\begin{figure}[t]
\centering
\includegraphics[width=0.8\linewidth]{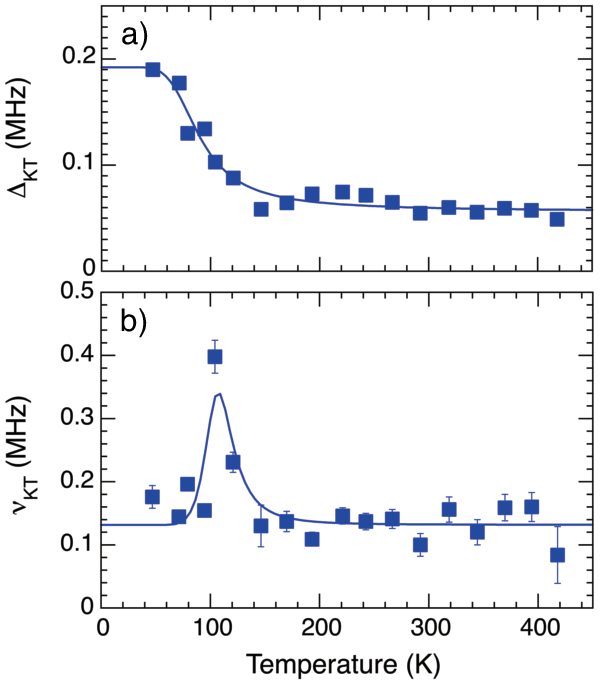}
\caption{Temperature dependence of (a) the linewidth $\Delta_{\rm KT}$ and (b) fluctuation frequency   $\nu_{\rm KT}$ deduced from the $\mu$SR time spectra observed in FAPbI$_3$ (after Ref.~\cite{Hiraishi:23}). The solid lines are the best fits with Eqs.~(2)--(4) in the main text (see Table \ref{tabfpi} for the parameter values).} 
\label{figfpi}
\end{figure}

The solid lines in Fig.~\ref{figfpi} show the best fits for these parameters using Eqs. (2)--(4) in the main text, where a temperature-independent offset $\nu_*$ was considered in Eq.~(3) to reproduce the data (whose origin is yet to be clarified).  As shown in Table \ref{tabfpi}, the values of $\nu_{\rm i0}$ and $E_{\rm i}$ obtained from these analyses are in reasonable agreement with each other, although the narrow temperature range of the peak in $\nu_{\rm KT}$ leads to relatively large uncertainty for the deduced parameter values. The parameters reported in Ref.\cite{Ito:24} are also quoted in Table \ref{tabfpi}, which are in excellent agreement with those deduced using the empirical model. Based on these results, it can be concluded that the main features of $\mu$SR data observed in FAPbI$_3$ have come to a unified understanding.

\begin{table}[b]
\begin{tabular}{c|cccc}
\hline\hline
\multirow{2}{*}{$\Delta_{\rm KT}$} & $\Delta$ (MHz) & $Q$ & $\nu_{\rm i0}$ (s$^{-1}$) & $E_{\rm i}$ (eV) \\
\cline{2-5}
&  0.192(11) & 0.92(23)& $5(5)\times10^{8}$ & 0.04(1)  \\
\hline
Ref.\cite{Ito:24} &  0.191(1) & 0.947(3) & $3.6(6)\times10^{8}$ & 0.054(2) \\
\hline\hline
\end{tabular}

\begin{tabular}{c|cccccc}
\hline\hline
\multirow{2}{*}{$\nu_{\rm KT}$} & $\Delta_{\rm p}$ (MHz) & $\nu_{\mu0}$ (MHz) & $E_\mu$ (eV) & $\nu_*$ (MHz) &  $\nu_{\rm i0}$ (s$^{-1}$) & $E_{\rm i}$ (eV) \\
\cline{2-7}
& 0.297(52)& -- & -- &0.13(1) & $7(6)\times10^{9}$ & 0.09(5)  \\
\hline\hline
\end{tabular}
\caption{Fitting parameters for muon and hydride ion jump motion in FAPbI$_3$  obtained from curve fits of $\Delta_{\rm KT}$ and $\nu_{\rm KT}$ versus temperature using Eqs.~(2)--(4), respectively. It is inferred from $\nu_{\rm KT}$ that muons are virtually immobile in the measured temperature range.}\label{tabfpi}
\end{table}

In contrast to quenched FAPbI$_3$, the KT parameters of MAPbI$_3$ and unquenched FAPbI$_3$ exhibit complex temperature dependence near the structural phase transition \cite{Koda:22,Hiraishi:23}. This is presumed to result from resonant correlations between the vibrational motion of the PbI$_3$ lattice and cation molecular rotational motion (which drives the structural transition, as evidenced by the distinct rise in the structural transition temperature by the deuteration of cation molecules \cite{Koda:22}), suggesting that the effective activation energy for rotational motion has changed as a result.

\subsection{Hyderide diffusion in LaH$_{3-2x}$O$_x$ ($x=0.25$)}

\begin{figure}[b]
\centering
\includegraphics[width=0.7\linewidth]{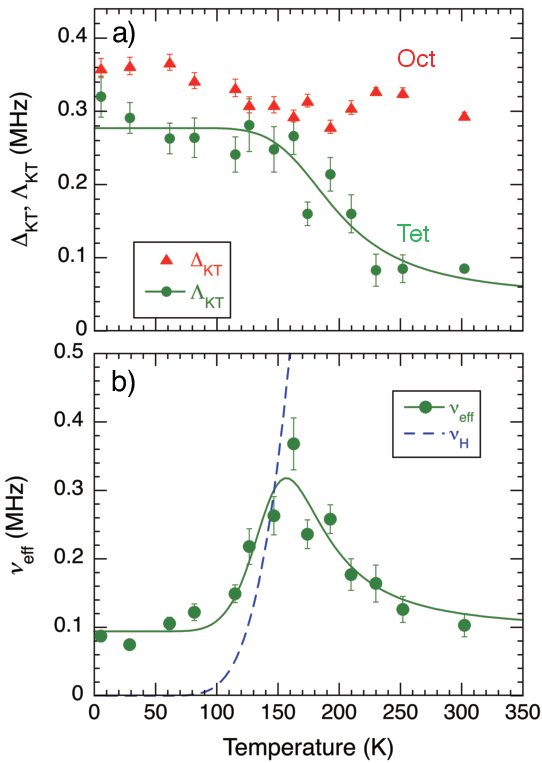}
\caption{Temperature dependence of the KT function parameters deduced by fits of the $\mu$SR time spectra in LaH$_{2.5}$O$_{0.25}$: (a) The linewidth for the static GKT function ($\Delta_{\rm KT}$) and for the dynamical LKT function ($\Lambda_{\rm KT}$). (b) The fluctuation rate $\nu_{\rm eff}$ ($\approx\nu_{\rm KT}$) of the internal magnetic field for the dynamical LKT function. The solid lines are the best fit with Eqs.~(2)--(4) in the main text. Dashed line in (b) represents jump frequency of H$^-$ ($\nu_{\rm H}=\nu_{\rm i}$) following the Arrhenius equation with the parameters in Table \ref{tablho}.} 
\label{figlho}
\end{figure}

Lanthanum hydroxide LaH$_{3-2x}$O$_x$, obtained by partially substituting H$^-$ in LaH$_3$ with O$^{2-}$, is known to exhibit extremely high hydride (H$^-$) ion conductivity at moderate to high temperatures \cite{Fukui:19,Fukui:22}. Recently, to gain insight into the microscopic mechanism of hydride conduction in this compound, $\mu$SR experiments were conducted on a sample with $x = 0.25$ at room temperature and below \cite{Hiraishi:26}. The previous neutron diffraction study inferred that the coordination of O atoms in this material is disordered due to strong Coulomb interactions with La, suggesting a glassy behavior \cite{Fukui:19}. Consequently, unlike previous examples of ion diffusion, the observed $\mu$SR time spectra exhibit a lineshape different from the Gaussian Kubo-Toyabe (GKT) relaxation function. More specifically, these are well described by the sum of Gaussian and Lorentzian Kubo-Toyabe (LKT) functions; it has been shown that the former originates from Mu$^-$ occupying octahedral sites, while the latter originates from Mu$^-$ occupying tetrahedral sites. Figure \ref{figlho} shows plots of the linewidths $\Delta_{\rm KT}$ (GKT), $\Lambda_{\rm KT}$ (LKT), and the fluctuation rate $\nu_{\rm eff}$ of $\Lambda_{\rm KT}$ versus $T$ reproduced from Figs.~3(b) and 3(c) in Ref. [\onlinecite{Hiraishi:26}].

\begin{table}[b]
\begin{tabular}{c|cccc}
\hline\hline
\multirow{2}{*}{$\Lambda_{\rm KT}$} & $\Delta$ (MHz) & $Q$ & $\nu_{\rm i0}$ (s$^{-1}$) & $E_{\rm i}$ (eV) \\
\cline{2-5}
&  0.278(13) & 0.97(11)& $4(14)\times10^{9}$ & 0.11(6)  \\
\hline
Ref.\cite{Hiraishi:26} &  0.277(14) & $\approx$1& -- & 0.11(3) \\
\hline\hline
\end{tabular}

\begin{tabular}{c|cccccc}
\hline\hline
\multirow{2}{*}{$\nu_{\rm eff}$} & $\Delta_{\rm p}$ (MHz) & $\nu_{\mu0}$ (MHz) & $E_\mu$ (eV) & $\nu_*$ (MHz) &  $\nu_{\rm i0}$ (s$^{-1}$) & $E_{\rm i}$ (eV) \\
\cline{2-7}
& 0.317(36)& -- & -- &0.094(15) & $2(2)\times10^{8}$ & 0.08(2)  \\
\hline\hline
\end{tabular}
\caption{The fitting parameters in Eqs.~(2)--(4) for H$^-$/Mu$^-$ ion jump motion in LaH$_{2.5}$O$_{0.25}$ obtained from curve fits of $\Lambda_{\rm KT}$ and $\nu_{\rm eff}$ versus temperature, respectively. It is inferred from $\nu_{\rm eff}$ that Mu$^-$ is virtually immobile in the measured temperature range.}\label{tablho}
\end{table}

The GKT component is quasistatic at all measured temperatures, and fluctuations in the internal magnetic field are observed only in the LKT component. Considering that H$^-$/Mu$^-$ at the T site is energetically more stable than that at the O site, this result suggests that H$^-$/Mu$^-$ diffusion is primarily due to jumps between T sites via vacant O sites. While the temperature dependence of the linewidth $\Lambda_{\rm KT}$ was analyzed using a two-state model \cite{Kadono:26} to yield the activation energy of 0.11(3) eV, it was difficult to quantitatively interpret the temperature dependence of $\nu_{\rm eff}\approx\nu_{\rm KT}$ which peaks around 160 K. 

The solid lines in Fig.~\ref{figlho} show the best fits for the LKT function parameters using Eqs. (2)--(4) in the main text, where a temperature-independent offset $\nu_*$ was considered again in Eq.~(3). 
 As shown in Table \ref{tablho}, the values of $\nu_{\rm i0}$ and $E_{\rm i}$ obtained from these analyses are in good agreement with each other, corroborating the validity of this model. In particular, it was reconfirmed that $E_{\rm i}$ agrees with the activation energy of the T-O jump estimated from neutron quasi-elastic scattering experiments. On the other hand, it also became clear that the prefactor $\nu_{\rm i0}$ is considerably smaller than the Debye frequency. This may be due to the anharmonic character of the potential energy experienced by H$^-$/Mu$^-$ ions (see below).
\begin{figure}[t]
\centering
\includegraphics[width=0.8\linewidth]{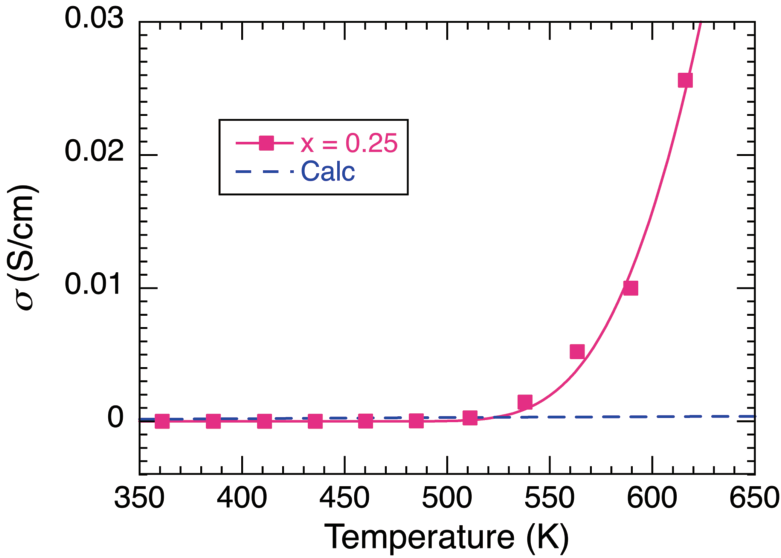}
\caption{Temperature dependence of ionic conductivity $\sigma$ in LaH$_{2.5}$O$_{0.25}$ (solid squares: after Ref.~\cite{Fukui:19}). Dashed line shows $\sigma$ predicted from the pre-factor and activation energy deduced from $\mu$SR results, and the solid line represents the Vogel-Fulcher-Tammann relation describing glass transition \cite{Hiraishi:26}.  } 
\label{figsgm}
\end{figure}

Because we were able to evaluate $\nu_{\rm i0}$ through these analyses, we can use the following equation to estimate the diffusion coefficient--and, by extension, ion conductance--from the jump frequency \cite{Habasaki:17}:
\begin{equation}
\sigma=\frac{nq^2}{2kT}\frac{d^2_{\rm TO}\nu_{\rm i0}}{z}\exp\left[-\frac{E_{\rm i}}{kT}\right],\label{sgmu}\\
\end{equation}
where $n$ is the ion density ($=2.2\times10^{22}$ cm$^{-3}$ for $x=0.25$), $q$ is the electric charge ($=e^-$), $d_{\rm TO}$ is the distance between T and O sites ($=3.75$ \AA), and $z$ is the number of the nearest neighboring O sites ($=4$); the H$^-$ diffusion is known to be dominated by the over-barrier jump between T sites via O sites.   Using $\nu_{\rm i0}=2\times10^8$ s$^{-1}$ and $E_{\rm i}=0.1$ eV, the temperature dependence of $\sigma$ predicted by Eq.~(\ref{sgmu}) is plotted in Fig.~\ref{figsgm}. The actually observed ionic conductivity is slightly lower than predictions at temperatures of $\approx$500 K and below, but far exceeds beyond those temperatures. This supports the view that the microscopic mechanism of fast diffusion at $T>500$ K is due to fluidic ion transport (which is subject to freezing due to glass transition around $T_{\rm g}\approx500$ K and below) rather than the classical over-barrier jump mechanism \cite{Hiraishi:26}.

\subsection{Li/Nb diffusion in Li$_y$Nb$_x$Ti$_{1-x}$O$_2$ and Nb$_x$Ti$_{1-x}$O$_2$}

In recent years, the demand for high-rate and fast-charging lithium-ion batteries has prompted various chemical substitution strategies to improve anode performance, such as doping anatase titanium dioxide (TiO$_2$) with pentavalent niobium. While conventional understanding generally posits that alkali ions like lithium are the sole mobile species while the transition metal framework remains structurally static, a recent study by Forslund {\it et al.}~\cite{Forslund:25} has reported contrasting microscopic evidence based on $\mu$SR experiment. 

The authors performed ZF/LF-$\mu$SR measurements on Li$_y$Nb$_x$Ti$_{1-x}$O$_2$ samples with $x=0.1$ and $0.01$ respectively for $y=0$ and $1$, and analyzed $\mu$SR time spectra by curve fits using the standard Kubo-Toyabe function $G_z^{\rm KT}(t)$ to deduce the linewidth $\Delta_{\rm KT}$ and its fluctuation rate $\nu_{\rm KT}$.  Following previous studies of this kind, they interpret the temperature ($T$) dependence of $\nu_{\rm KT}$ as that due to the diffusive motion of cations. Specifically, they assume that Nb ions exhibit jump motion with a frequency observable with $\mu$SR in the samples with $y=0$, and they discuss the activation energy of Nb diffusion by interpreting specific temperature regions of $\nu_{\rm KT}(T)$ using the Arrhenius law.  Based on these interpretations, the authors  suggest that transition metal mobility may be more prevalent in substituted battery materials than previously assumed, proposing that such dynamics could play a role in the electrochemical behavior of these anodes.

\begin{figure}[t]
\centering
\includegraphics[width=0.8\linewidth]{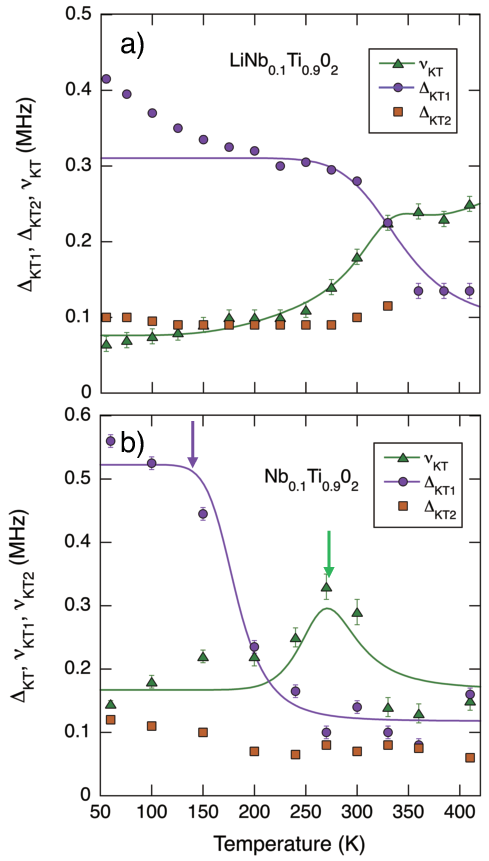}
\caption{The Kubo-Toyabe linewidth and fluctuation frequency versus temperature in (a) LiNb$_{0.1}$Ti$_{0.9}$O$_2$ and (b) Nb$_{0.1}$Ti$_{0.9}$O$_2$ quoted from Ref.~[\onlinecite{Forslund:25}]. The arrows in (b) are the characteristic temperature $T^*$ for $\Delta_{\rm KT1}$ and $\nu_{\rm KT}$. Solid lines are the best fits of $\Delta_{\rm KT1}$ with Eq.~(2), and those of $\nu_{\rm KT}$ with Eqs.~(3) and (4), respectively. (See Table \ref{tabnto} for the parameter values deduced from fits.) Note that the fit for $\nu_{\rm KT}$ in (b) is so poor ($R=0.701$) that the deduced parameters are totally unreliable.}
\label{fignto}
\end{figure}

Here, let us use our empirical model to examine their interpretation using the examples with $x=0.1$, LiNb$_{0.1}$Ti$_{0.9}$O$_2$ (LNTO) and Nb$_{0.1}$Ti$_{0.9}$O$_2$ (NTO). Figure \ref{fignto} reproduces the plots of $\Delta_{\rm KT1}$, $\Delta_{\rm KT2}$, and $\nu_{\rm KT}$  shown in Figs.~2(A) and (C) of Ref.[\onlinecite{Forslund:25}]. First, considering the case of LNTO (Fig.~\ref{fignto}(a)), $\Delta_{\rm KT1}$ among the two linewidths begins to decrease at temperatures above $T^*\approx300$ K, and $\nu_{\rm KT}$ exhibits a peak near $T^*$. These temperature dependencies are consistent with the behavior predicted by our empirical model; in particular, the latter characteristic strongly suggests that the fluctuations in $\Delta_{\rm KT1}$ are caused by the motion of cations. 

The results of curve fits for $\Delta_{\rm KT1}(T)$ and $\nu_{\rm KT}(T)$ using Eqs. (2)--(4) in the main text are shown as solid lines in Fig.~\ref{fignto}(a); the $\Delta_{\rm KT1}$ fit was performed using only data for $T>200$ K, and a temperature-independent offset $\nu_*$ was considered in Eq.~(3). The parameter values obtained from the fit are shown in Table \ref{tabnto}. The values of $\nu_{\rm i0}$ and $E_{\rm i}$ obtained respectively from $\Delta_{\rm KT1}$ and $\nu_{\rm KT}$ agree with each other within the margin of error, providing evidence that the interpretation based on this model is valid. Furthermore, these values are consistent with that known for the Li ion diffusion in other battery materials, where $\nu_{\rm i0}\approx10^{10}$--$10^{14}$ s$^{-1}$ and $E_{\rm i}\approx0.15$--0.7 eV \cite{Ito:26}.

Table \ref{tabnto} also quotes the values of $\nu_{\rm i0}$ and $E_{\rm i}$ from Ref.\cite{Forslund:25}, where $\nu_{\rm KT}$ at temperatures lower than $T^*$ was analyzed using the Arrhenius equation. The values of $\nu_{\rm i0}$ are lower than those obtained by our empirical model by several orders of magnitude, and $E_{\rm i}$ are underestimated by about one-sixth; these trends are in line with the results of benchmarks performed using simulation data.

\begin{table}[t]
\begin{tabular}{c|c|ccccc}
\hline\hline
\multirow{3}{*}{$\Delta_{\rm KT}$}& $y$  & $\Delta$ (MHz) & $Q$ & $\nu_{\rm i0}$ (s$^{-1}$) & $E_{\rm i}$ (eV) \\
\cline{2-6}
&$1$ & 0.31(1) & 0.915(45)& $4(6)\times10^{11}$ & 0.33(5)  \\
&$0$& 0.522(7) & $0.92(12)$ & 6(8)$\times10^{11}$ & 0.18(2)  \\
\hline\hline
\end{tabular}

\begin{tabular}{c|c|cccccc}
\hline\hline
\multirow{3}{*}{$\nu_{\rm KT}$} &$y$ & $\Delta_{\rm p}$ (MHz) & $\nu_{\mu0}$ (MHz) & $E_\mu$ (eV) & $\nu_{\rm i0}$ (s$^{-1}$) & $E_{\rm i}$ (eV) \\
\cline{2-7}
&$1$& 0.08(2)& 1.2(4) & 0.073(10) & $2(7)\times10^{10}$ & 0.34(13)  \\
&$0$ & 0.08(2) & -- & -- & $2\times10^{10}$ & 0.260(1) \\
\hline
\multirow{2}{*}{Ref.\cite{Forslund:25}}& $1$& --& -- & -- & $1.3(2)\times10^{6}$ & 0.052(4)  \\
& $0$ & -- & -- & -- & $5(1)\times10^{5}$ & 0.013(2) \\
\hline\hline
\end{tabular}
\caption{Fitting parameters for muon and Li-ion jump motion in LiNb$_{0.1}$Ti$_{0.9}$O$_2$ ($y=1$) and Nb$_{0.1}$Ti$_{0.9}$O$_2$ ($y=0$) obtained from curve fits of $\Delta_{\rm KT}$ and $\nu_{\rm KT}$ versus temperature using Eqs.~(2)--(4), respectively. $\nu_*$ is 0.076(4) for $y=1$ and 0.17(2) for $y=0$. }\label{tabnto}
\end{table}

In contrast, the fact that our empirical model does not yield consistent analytical results for the NTO case suggests that the observed $\Delta_{\rm KT}$ and $\nu_{\rm KT}$ do not contain information regarding the diffusion of cations.  This strongly suggests that, within the temperature range over which the measurements were conducted, the Nb jump frequency has not reached the $\mu$SR sensitivity region, and is consistent with the results of recent DFT calculations indicating that the activation energy for Nb diffusion via cation vacancies is 1.85 eV \cite{Ito:pc}, which is also comparable with the activation energy for the cation self-diffusion in oxides \cite{Lindner:55}.  Considering these results in NTO, we can conclude that the cation motion suggested in LNTO is attributable to that of Li ions.

\end{document}